\documentclass[runningheads,a4paper]{svmult}

\usepackage{makeidx}   % allows index generation
\usepackage{graphicx}  % standard LaTeX graphics tool
\usepackage{subeqnar}  % subnumbers individual equations
\usepackage{multicol}  % used for the two-column index
\usepackage{physprbb}  % modified textarea for proceedings,
\makeindex             % used for the subject index
\begin{document}
\title*{A GRB Detection System using 
the BGO-Shield of the INTEGRAL-Spectrometer SPI}
\toctitle{A GRB Detection System using 
the BGO-Shield of the INTEGRAL-Spectrometer SPI}
% allows explicit linebreak for the table of content
%
%
\titlerunning{SPI/ACS GRB Detection System}
% allows abbreviation of title, if the full title is too long
% to fit in the running head
%
\author{Andreas von Kienlin
\and Nikolas Arend
\and Giselher G. Lichti}
\authorrunning{Andreas von Kienlin et al.}
% if there are more than two authors,
% please abbreviate author list for running head
%
%
\institute{Max-Planck-Institut f\"ur extraterrestrische Physik, 
85741 Garching, Germany}

\maketitle              % typesets the title of the contribution

\begin{abstract}
The anticoincidence shield (ACS) of the INTEGRAL-spectrometer SPI consists of 
512 kg of BGO crystals. This massive scintillator allows the measurement of
gamma-ray bursts (GRBs) with a very high sensitivity. Estimations have 
shown that with the ACS some hundred gamma-ray bursts per year on the 
5 $\sigma$ level can be detected, having an equivalent sensitivity to 
BATSE. The GRB detection will be part of the real-time INTEGRAL 
burst-alert system (IBAS). The ACS branch of IBAS will produce burst 
alerts and light curves with 50 ms resolution. It is planned to use ACS 
burst alerts in the 4th interplanetary network (IPN) \cite{hurley}.  
\end{abstract}

\section{The Anticoincidence Shield of SPI}
\begin{figure}[t]
\begin{center}
\includegraphics[width=.75\textwidth]{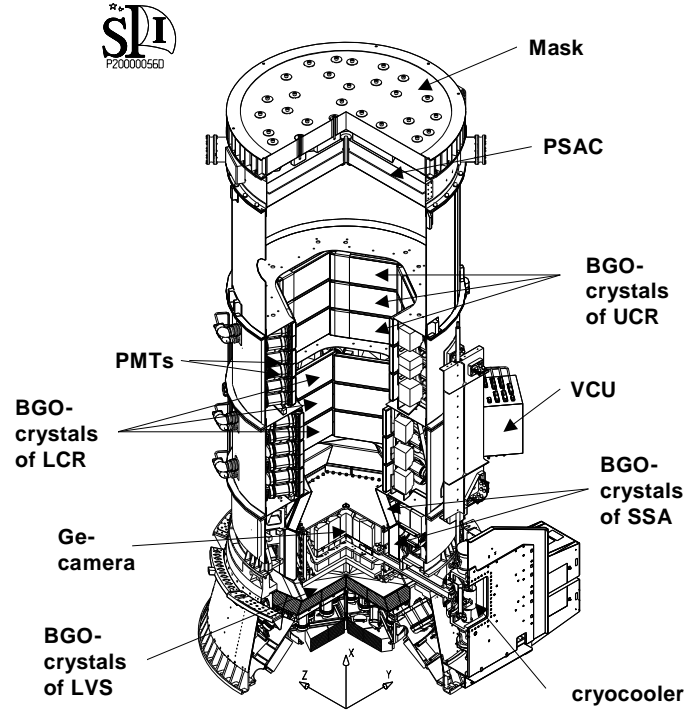}
\end{center}
\caption[]{INTEGRAL Spectrometer SPI}
\label{spi}
\end{figure}

The spectrometer SPI \cite{vedrenne} is one of the two main instruments on INTEGRAL, 
one of ESA's next missions devoted to $\gamma$-ray research.
Fig. \ref{spi} shows a drawing of the spectrometer SPI.  
The camera of SPI, which consists of 19 cooled high-purity germanium detectors, 
is shielded on the side walls and rear side by a large anticoincidence shield (ACS). 
The field of view of the camera is defined by the upper opening of the ACS.  The imaging 
capability of the instrument is attained by a passive-coded mask on the top. Below 
the mask a plasticscintillator anticoincidence (PSAC) takes care for the reduction 
of the 511 keV background, which is mainly generated by particle interactions in the mask.

The ACS  consists of 91 BGO crystals which are arranged in 4 subunits. The units of the upper
veto shield (UVS) consists of the upper collimator ring (UCR), the lower collimator ring (LCR) and 
the side shield assembly (SSA), each containing 18 crystals which are arranged hexagonally around the 
cylindrical axis of SPI. The lower veto shield (LVS), consisting of 36 crystals, is assembled 
as a hexagonal shell. The thickness of the crystals increases from 16 mm at the top (UCR) 
to 50 mm at the bottom (LVS). The total mass of BGO used for the ACS is 512 kg resulting in the
obvious use of the ACS as a burst monitor.  

Each BGO crystal of the ACS (with one exception) is viewed by two photomultipliers (PMTs). 
Due to the redundancy concept used for the ACS, each of the 91 front-end electronic boxes 
(FEEs) sums the anode signals of two PMTs, which are viewing different BGO crystals 
(in most cases neighbouring crystals). This cross strapping of FEEs and BGO 
units leads in a failure case of one single PMT or FEE not to the loss of a complete BGO crystal. 
It emerges that a disadvantage of this method is an uncertainty in the energy-threshold value of 
individual FEEs, caused by a different light yield of neighbouring BGO-crystals and different PMT
properties like quantum efficiency and amplification. A result of this is that the threshold 
extends over a wide energy range and is not at all sharp. 
The energy-threshold settings of the ACS depend on a tradeoff between 
background reduction and deadtime for the SPI camera.

\section{The ACS as GRB Monitor}
The main task of the ACS as a detector is the veto generation for charged particles and
$\gamma$-rays coming from outside the FoV. But there are also data
which can be used for scientific purposes. The ACS housekeeping (HK) data include 
the values of the overall veto counter of the veto control unit (VCU) and 
the individual ratemeter values of each FEE. Both HK data are suitable for burst detection.
The count rate of the overall veto counter (ORed veto signals of all 91 FEEs) is sampled every
50 ms. A packet, containing 160 consecutive count rates, will be transmitted every 8 sec
to ground. If no gap in the telemetry stream occurs one could have a continuous ACS veto-rate 
light curve with 50 ms binning. The measurement time of the individual FEE ratemeter can be 
adjusted between 0.1 and 2 sec. All 91 FEEs are read out successively in groups of 8 FEEs every 
8 sec. The read out of all 91 ratemeter values thus needs 96 sec. In distinction to the VCU overall veto counter
the individual ratemeter values do not yield a continuous stream. Additionally the
values of different FEE groups are shifted by a time interval of 8 sec.
So it is very difficult to derive the burst-arrival direction from these individual counting rates.

\section{SPI/ACS Burst-Alert System}
\begin{figure}[t]
\begin{center}
\includegraphics[width=.6\textwidth, angle=270]{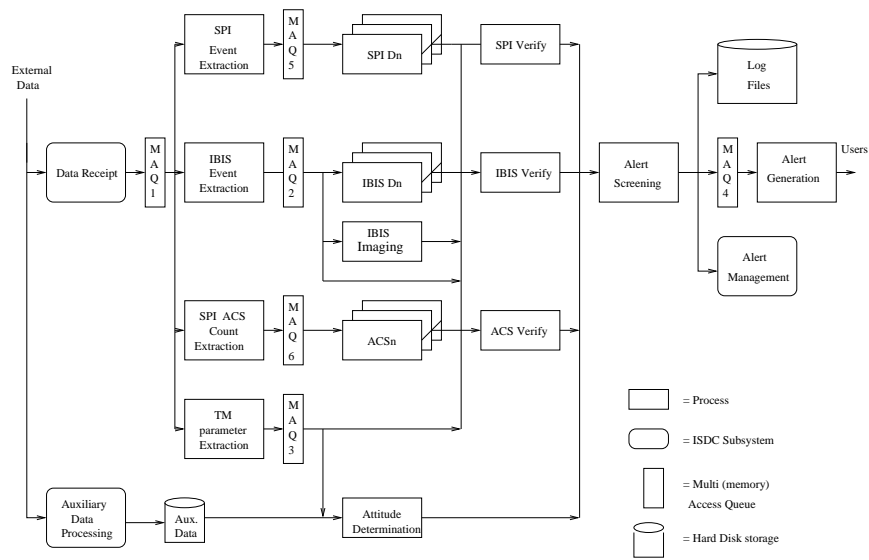}
\end{center}
\caption[]{IBAS structure}
\label{ibas}
\end{figure}

The search for GRBs in the ACS veto-rate light curve  and the generation of alerts 
will be performed automatically on ground at the INTEGRAL Science Data Center (ISDC).
The ISDC system responsible for the monitoring of the data for burst occurence of all 
INTEGRAL instruments is the  INTEGRAL Burst Alert System (IBAS) \cite{sandro}. The structure 
of IBAS is shown in Fig.\ref{ibas}. 
The SPI/ACS Burst Alert System (SACS-BAS) is one branch of IBAS. The telemetry (TM) files of 
the INTEGRAL satellite are transmitted via the Mission Operation Center (MOC) to the ISDC and are
then directly fed into the real-time telemetry (RTTM) receiver of the IBAS system. After 
distribution and extraction of the relevant data, each IBAS branch is processing a burst-search 
algorithm with a subsequent verify procedure. The trigger algorithm used for SACS-BAS is looking 
for a significant excess with respect to a running average, comparable to the trigger algorithm 
used for other spacecrafts (e.g. ULYSSES). Due to the sufficient computing power available on 
ground it is possible to run several burst-search and burst-verify processes in parallel.
SACS-BAS has also implemented this option: several trigger processes with different time bin 
durations will run in parallel in order to be able to trigger on bursts with different temporal
behaviour; several verify processes with different criteria, gain for the supression of fault triggers 
generated by backgound variations, could be started in SACS-BAS after a burst alert.
Up to now SACS-BAS is only reading the overall-veto counter values. But it is planned also to include the
read out of rate meter-values of individual FEEs into the SACS-BAS routine. This will allow a rough 
estimation of the GRB arrival direction. An accuracy of about $10^{o}$ - $20^{o}$ will be 
enough to distinguish between the two arrival-cone intersections of the interplanetary network (IPN). 
The SACS-BAS trigger algorithm will be tested with generated TM data of simulated 
burst data plus background. After launch all parameters for the trigger algorithm
and verify criteria will be optimised.

The output of SACS-BAS will be burst alerts, containing information about the trigger time in 
universal time (UT), the spacecraft position and its attitude. 
The alerts will be transmitted to subscribed users by e-mail and/or direct TCP/IP socket.
Especially for the IPN the burst time history ($\sim 100$~s) is important for the alignment
of the light curves obtained from different spacecrafts. For this purpose the time history togther 
with the pre-trigger time-history  ($\sim 5$~s) will be transmitted to the IPN. It is important 
for the IPN to know the burst arrival time with a millisecond accuracy. 
As already shown in \cite{grl} this is possible for the ACS overall counter values.

\section{Sensitivity Estimation}
An estimation of the expected rate of GRBs detected by ACS has already been given in \cite{grl}.
For an effective area of about 3000~cm$^2$, an ACS background rate between $\sim 80000$~cts/s
and $\sim 160000$~cts/s, an ACS threshold of 80 keV and a time binning of 50 ms the minimal detectable 
energy flux lies between $2 \leftrightarrow 2.8 \times 10^{-6} \frac{{\rm erg}}{{\rm cm}^2{\rm sec}}$.
For a time binning of 1 sec the sensitivity increases to $5 \times 10^{-7} \frac{{\rm erg}}{{\rm cm}^2{\rm sec}}$.
Using the logN-logP distribution, measured by BATSE and PVO \cite{fen} one can derive the number of bursts
which will be observed with the ACS in one year.
The resulting values are $\sim 50$ bursts/year for 50 ms integration time and $\sim 280$ bursts/year for
1 sec integration time.
The response of ACS depends on the infalling direction of a burst due to projection effects of other
ACS crystals and due to shielding of neighbouring instruments and spacecraft structure.
The burst intensity could be determined once the infalling direction is known. This is possible only via 
Monte-Carlo simulations using the INTEGRAL mass model. Similar simulations have been performed by P. Jean
et al. \cite{pjean} in order to determine the sensitivity of the ACS for detection of novae.

%INDEX%%%%%%%%%%%%%%%%%%%%%%%%%%%%%%%%%%%%%%%%%%%%%%%%%%%%%%%%%%%%%%%
% Please check with the editor of your book whether he plans to
% include a "mutual" subject index - if so, please code your entries
% in the standard syntax. For your own purposes you may print your
% "personal" index by using the following commands:
%
%\clearpage
%\addcontentsline{toc}{section}{Index}
%\flushbottom
%\printindex
%%%%%%%%%%%%%%%%%%%%%%%%%%%%%%%%%%%%%%%%%%%%%%%%%%%%%%%%%%%%%%%%%%%%%

\end{document}